\documentstyle[12pt]{article}
\newtheorem{Lemma}{Lemma}
\newtheorem{Proposition}{Proposition}
\newtheorem{Theorem}{Theorem}
\newtheorem{Corollary}{Corollary}
\def\proof{\par{\it Proof}. \ignorespaces}
\def\endproof{{\ \vbox{\hrule\hbox{%
   \vrule height1.3ex\hskip0.8ex\vrule}\hrule }}\par}
\newenvironment{Proof}{\proof}{\endproof}

\begin{document}

\title{ Toda Hierarchy with Indefinite Metric}

\author{Yuji KODAMA\thanks{\it e-mail:\ kodama@math.ohio-state.edu } \ \
 and  \ Jian YE\thanks{\it e-mail:\ ye@math.ohio-state.edu}
 \\{\it Department of Mathematics,
Ohio State University,}
\\ {\it Columbus, OH 43210, USA} \\ }

\date{May 15, 1995}

\maketitle

\begin{abstract}

We consider a generalization of the full symmetric Toda hierarchy
where the matrix $\tilde {L}$ of the Lax pair is given by
$\tilde {L}=LS$,
with a full symmetric matrix $L$ and a nondegenerate diagonal matrix $S$.
The key feature of the hierarchy is that the inverse scattering data
includes a class of noncompact groups of matrices,
such as $O(p,q)$.
We give an explicit formula for the solution to the initial value problem
of this hierarchy.  The formula is obtained by generalizing
the orthogonalization procedure of Szeg\"{o}, or
the QR factorization method of Symes.
The behaviors of the solutions are also studied. Generically, there
are two types of solutions, having either sorting property or blowing up
to infinity in finite time.
The $\tau$-function structure for the tridiagonal hierarchy is also studied.
\end{abstract}

\medskip


\vskip 3cm

\par\bigskip\bigskip


\section{Introduction}
\renewcommand{\theequation}{1.\arabic{equation}}\setcounter{equation}{0}

The finite non-periodic Toda lattice hierarchy can be
written in the Lax form \cite{flash}, \cite{moser} with variables
${\bf t} := (t_1,t_2,\cdots)$,
\begin{eqnarray}
\label{toda}
\frac{\partial }{\partial t_{n}} L = \left[ B_{n} \ , \ L \right] \ ,
\ n = 1, 2, \cdots
\end{eqnarray}
where $L$ is an $N \times N$ symmetric ``tridiagonal'' matrix with real
entries,
\begin{eqnarray}
\label{l}
L = \left(
\begin{array}{lllll}
a_1 & b_1 & 0 & \ldots & 0 \\
b_1 & a_2 & b_2 & \ldots & 0 \\
0 & \vdots & \ddots & \vdots & 0 \\
0 & \ldots & \ldots & a_{N-1} & b_{N-1} \\
0 & \ldots & \ldots & b_{N-1} & a_N \\
\end{array}
\right)
\end{eqnarray}
and $B_{n}$ is the
skew symmetric matrix defined by
\begin{eqnarray}
\label{skew}
B_{n} = \prod_{a} L^{n} := \left( L^{n} \right)_{>0} -
\left( L^{n} \right)_{<0} \ .
\end{eqnarray}
Here $\left( L^{n} \right)_{>0 \ (<0)}$ denotes the strictly upper (lower)
triangular part of $L^{n}$.  (We formally write an infinite number of flows
in (\ref{toda}), even though there are at most $N$ independent flows.)
In particular, (\ref{toda}) for n=1 with $t := t_1$ has the form
\begin{eqnarray}
\label{exa}
\frac{\partial a_k}{\partial t}=2(b_k^2-b_{k-1}^2),
\end{eqnarray}
\begin{eqnarray}
\label{exb}
\frac{\partial b_k}{\partial t}=b_k(a_{k+1}-a_k).
\end{eqnarray}
This system describes a hamiltonian system of $N$ particles on a line
interacting pairwise with exponential forces.
The hamiltonian for the system
is given by
\begin{eqnarray}
\label{ham}
H=\frac{1}{2}\sum_{k=1}^{N}y_k^2 + \sum_{k=1}^{N-1}\exp(x_k-x_{k+1}),
\end{eqnarray}
where the canonical variables $(x_k, y_k)$ are related to $(a_k, b_k)$
by
\begin{eqnarray}
\label{a}
a_k=-\frac{y_k}{2},
\end{eqnarray}
and
\begin{eqnarray}
\label{b}
b_k=\frac{1}{2}\exp \left(\frac{x_k-x_{k+1}}{2} \right).
\end{eqnarray}

\medskip

The $\tau$-functions
were introduced in \cite{naka} to study the
Toda equation (\ref{exa}) and (\ref{exb}).
Writing $a_i$'s with the $\tau$-function $\tau_i$'s and $t=t_1$,
\begin{eqnarray}
\label{atau}
a_i=\frac{1}{2}\frac{\partial}{\partial t}\log\frac{\tau_i}{\tau_{i-1}},
\end{eqnarray}
$b_i$ can be expressed as
\begin{eqnarray}
\label{btau}
b_i^2=\frac{1}{4}\frac{\tau_{i+1}\tau_{i-1}}{\tau_i^2}.
\end{eqnarray}
Then (\ref{exa}) and (\ref{exb}) become
\begin{eqnarray}
\label{etau}
\frac{1}{4}\frac{\partial^2}{\partial
t^2}\log\tau_i=\frac{\tau_{i+1}\tau_{i-1}}{\tau_i^2}.
\end{eqnarray}
These $\tau$-functions $\tau_i$ have a simple structure, that is,
a symmetric wroskian given by \cite{nk},
\begin{eqnarray}
\label{tau}
\tau_i = \left|
\begin{array}{lllll}
g & g_1 & g_2 & \ldots & g_{i-1}\\
g_1 & g_2 & g_3 & \ldots &  g_i\\
\  & \vdots & \ddots & \vdots & \  \\
g_{i-1} & g_i & \ldots & \ldots & g_{2i-2}\\
\end{array}
\right|
\end{eqnarray}
where the entries $g_n$ are defined by $g_n:=(1/2)\partial g/\partial t_n$,
with a function $g$ satisfying the linear equations
\begin{eqnarray}
\label{geq}
\frac{1}{2}\frac{\partial g}{\partial t_n}\ =
\ \frac{1}{2^n}\frac{\partial^n g}{\partial t^n}.
\end{eqnarray}
These $\tau$-fuctions are also shown to be positive definite in \cite{nk}. It
is a key property of their relation to the moment problem of Hamburger.
The equation (\ref{etau}) is also expressed by the well-known Hirota
bilinear form
\begin{eqnarray}
\label{hirota}
D_t^2\tau_i\cdot\tau_i=8\tau_{i+1}\tau_{i-1} \ ,
\end{eqnarray}
where the Hirota derivative is defined by
\begin{eqnarray}
\label{Derivative}
\nonumber
(D_tf\cdot g)(t)&:=&\frac{d}{ds}f(t+s)g(t-s)|_{s=0} \\
&=& \left({\frac{df}{dt}g-f\frac{dg}{dt}} \right)(t).
\end{eqnarray}

\medskip

There have been extensive studies on the hierarchy (\ref{toda}) and its
generalizations.
One of them is to extend $L$ from ``tridiagonal'' to ``full symmetric''. This
more general system, which we call ``full symmetric Toda hierarchy'', was
shown by Deift et. al. in \cite{DNLT} to be a completely integrable
hamiltonian system.
The inverse scattering scheme for (\ref{toda}) with $L$ being any symmetric
 matrix consists of two linear equations,
\begin{eqnarray}
\label{comp1}
L \Phi &=& \Phi \Lambda  \ , \\
\label{comp2}
\frac{ \partial }{\partial t_{n}} \Phi &=& B_{n} \Phi \ ,
\end{eqnarray}
where $\Phi$ is the orthogonal eigenmatrix of $L$, and $\Lambda$ is $diag
\left( \lambda_1, \cdots, \lambda_N \right)$. The hierarchy (\ref{toda})
then results as the compatibility of these equations with $\partial \Lambda
/ \partial t_n = 0$ (iso-spectral deformation).
In \cite{KM}, Kodama and McLaughlin solved the initial value problem of
 (\ref{comp1}) and (\ref{comp2})
using the ``orthonormalization method'', and derived an explicit formula
of the solution
in a determinant form. They also showed that the generic solution assumes
the ``sorting property''. Here the sorting property means that
$L(t_1) \rightarrow
\Lambda=diag\left( \lambda_1, \cdots, \lambda_N \right)$ as $t_1 \rightarrow
\infty$, with the eigenvalues being ordered by $\lambda_1>\lambda_2>\cdots>
\lambda_N$.

\medskip

The hierarchy we consider in this paper is for
$\tilde {L} = LS$,
where $L$ is full symmetric and $S$ is a constant diagonal matrix,
$S= diag \left( s_1, \cdots, s_N \right)$ with nonzero ``real'' entries.
All the entries of $\tilde L$ are assumed to be real, unless otherwise
stated.
The hierarchy
is also defined as the
Lax form (\ref{toda}) with (\ref{skew}).
In particular, for the case of tridiagonal $\tilde L$ with n=1, we have
\begin{eqnarray}
\label{exsa}
\frac{\partial a_k}{\partial t}=2(s_kb_k^2-s_{k-1}b_{k-1}^2),
\end{eqnarray}
\begin{eqnarray}
\label{exsb}
\frac{\partial b_k}{\partial t}=b_k(s_{k+1}a_{k+1}-s_ka_k).
\end{eqnarray}
In analogue to the nonperiodic Toda equation with (\ref{ham}),
the hamiltonain $\tilde H$ for the above equations is given by
\begin{eqnarray}
\label{ham1}
\tilde H=\frac{1}{2}\sum_{k=1}^{N}y_k^2 +
\sum_{k=1}^{N-1}s_ks_{k+1}\exp(x_k-x_{k+1}),
\end{eqnarray}
with the
change of variables
\begin{eqnarray}
\label{as}
s_ka_k=-\frac{y_k}{2},
\end{eqnarray}
and
\begin{eqnarray}
\label{bs}
s_ks_{k+1}b_k=\frac{1}{2}\exp\left({x_k-x_{k+1} \over 2}\right) \ .
\end{eqnarray}
Note from (\ref{ham1}) that $\tilde H$
includes some attractive forces, thereby is not positive definite.
One then expects a `` blowing up'' in solutions.

\medskip

We study the initial value problem of the hierarchy by the ``inverse
scattering method".
The inverse scattering scheme is
also given by (\ref{comp1}) and (\ref{comp2}), where the eigenmatrix $\Phi$ is
now normalized to satisfy
\begin{eqnarray}
\label{PhiS}
\Phi S^{-1} \Phi^T \ = S^{-1}, \ \ \Phi^T S \Phi \ = S .
\end{eqnarray}
In the case of the full symmetric Toda hierarchy, where
 $S$ is the identity matrix,
$\Phi$ is given by an orthogonal matrix, $\Phi \in O(N)$.
As a special case of (\ref{PhiS}), we have $\Phi \in O(p,q)$ for
$S = diag(1, \cdots , 1, -1, \cdots , -1)$, with  $p +q =N$.
{}From the first quation in
(\ref{PhiS}), we define an inner product where $S^{-1}$ gives an indefinite
metric for $S$ being not positive definite. For this reason, we call the
hierarchy
considered in this paper
the ``Toda hierarchy with indefinite
metric''.

\medskip

The content of this paper is as follows:  We start with a preliminary
in Section 2 to give some background information on the Toda hierarchy
and the inverse scattering scheme.
First we show the compatibility of the  flows in (\ref{toda}).
Then we show that the eigenmatrix of $\tilde L$ can be chosen to satisfy
(\ref{PhiS}), and also (\ref{PhiS}) is invariant under the flows generated
by (\ref{comp2}).

\medskip

In Section 3, we show that
the hierarchy can be solved using the orthonormalization
method with respect to the indefinite metric.
The solution of the hierarchy turns out to be given by
the same solution formula
in \cite{KM} with a little modification.

\medskip

In Section 4, we discuss the behaviors of the solutions.
Due to the indefiniteness of the metric defined by
$S^{-1}$, the solution has richer
behaviors than the full symmetric case. Generically, in addition to the
sorting property, there are solutions blowing up to infinity in finite
time. The behaviors of the solutions are characterized by $S$ and the initial
conditions, {\it i.e.}, the eigenvalues $\lambda_i, i=1, \cdots, N$ of
$\tilde L$ and the initial eigenmatrix $\Phi^0$. In this paper, we obtain
the following
main results: If $S$ is positive definite, then generic solutions have
the sorting property (Theorem 2). If some eigenvalues of $\tilde{L}$ are
not real, or $\Phi^0$ is not real, then generic solutions blow up to infinity
in finite time (Theorem 3).

\medskip

In Section 5, we illustrate these results with explicit examples.

\medskip

In Section 6, we study $\tau$-function structures of the $\tilde L$ hierarchy.
In the case of tridiagonal $\tilde L$,
one can also introduce $\tau$-functions in
the same form as the symmetric wroskians (\ref{tau}).
However, our $\tau$-functions
are no longer positive definite as a result of the indefinite metric in
our hierarchy.
We also find a ``bilinear identity" generating relations among $\tau_i$'s
of (\ref{tau}).  From the bilinear identity, we then derive a hierarchy
written in Hirota's bilinear form
for the $\tau$-functions including (\ref{hirota}) as its first member.

\section{Preliminary}
\renewcommand{\theequation}{2.\arabic{equation}}\setcounter{equation}{0}

The hierarchy considered in this paper is defined as
\begin{eqnarray}
\label{todals}
\frac{\partial }{\partial t_{n}} \tilde L
= \left[ \tilde B_{n} \ , \ \tilde L \right] \ ,
\ n = 1, 2, \cdots
\end{eqnarray}
where $\tilde L=LS$ with $L$, a full symmetric matrix, and $S$,
a nondegenerate diagonal matrix, i.e., a symmetrizable matrix in the sense of
Kac \cite{kac}, and
$\tilde B_{n} := \prod_a \tilde L^n= (\tilde L^n)_{>0}-(\tilde L^n)_{<0}$.
One should note that among the infinite number of flows in the
hierarchy, only the first $N$ flows are independent, and the higher order flows
are related to these $N$ flows through $P(\tilde L)=0$, the characteristic
polynomial of $\tilde L$.

\medskip

Let us first show the compatibility of flows in (\ref{todals}),
that is,
${\partial^2 \tilde L} / {\partial t_m \partial t_n}
={\partial^2 \tilde L} / {\partial t_n \
\partial t_m}$. This can be shown from the ``zero curvature'' condition, i.e.
\begin{eqnarray}
\label{curvature}
\frac{\partial \tilde B_n}{\partial t_m}-\frac{\partial \tilde B_m}
{\partial t_n}= [\tilde B_m\ , \tilde B_n]\ .
\end{eqnarray}
\begin{Proposition}
\label{commutivity}
The flows in (\ref{todals}) commute with each other.
\end{Proposition}
\begin{Proof}
It suffices to show the zero curvature condition (\ref{curvature}).
{}From (\ref{todals}), we have
\begin{eqnarray}
\label{llb}
\frac{\partial \tilde L^n}{\partial t_m}- \frac{\partial \tilde L^m}
{\partial t_n} =
[\tilde B_m, \tilde L^n]-[\tilde B_n, \tilde L^m] \ ,
\end{eqnarray}
which can be also written as
\begin{eqnarray}
\label{bb}
[\tilde B_m, \tilde L^n]-[\tilde B_n, \tilde L^m]
=[\tilde B_n, \tilde B_m]+[\tilde B_m+\tilde L^m, \tilde B_n+\tilde L^n].
\end{eqnarray}
Note that $\tilde B_m+\tilde L^m$ is an upper triangular matrix. Taking the
lower triangular
projection of (\ref{llb}), we have
\begin{eqnarray}
\label{llbpl}
\frac{\partial (\tilde L^n)_{<0}}{\partial t_m}- \frac{\partial
(\tilde L^m)_{<0}}{\partial t_n} = ([\tilde B_n, \tilde B_m])_{<0}.
\end{eqnarray}
Similarly,
\begin{eqnarray}
\label{llbpu}
\frac{\partial (\tilde L^n)_{>0}}{\partial t_m}- \frac{\partial
\tilde (L^m)_{>0}}{\partial t_n} = -([\tilde B_n, \tilde B_m])_{>0}.
\end{eqnarray}
Subtracting (\ref{llbpu}) from (\ref{llbpl}), we get
\begin{eqnarray}
\label{almost}
\frac{\partial \tilde B_n}{\partial t_m}-\frac{\partial \tilde B_m}
{\partial t_n}=
([\tilde B_m\ , \tilde B_n])_{>0}+([\tilde B_m\ , \tilde B_n])_{<0}.
\end{eqnarray}
Since $[\tilde B_m\ , \tilde B_n]$ is skew-symmetric, $diag([\tilde B_m\ ,
\tilde B_n])=0$. This completes the proof.
\end{Proof}

\medskip

{\bf Remark 1} Since the specific form of $\tilde L$ is not used in this proof,
Proposition 1 is valid for an arbitrary matrix.

\medskip

In the case of $L$ being full symmetric, $L$ can be
diagonalized by an orthogonal
matrix, and the flows defined by (\ref{toda}) are compatible
with the choice of skew-symmetric $B_n$ by (\ref{skew}).
To set up the inverse
scattering scheme for our system,
we need the following lemma from linear algebra:
\begin{Lemma}
\label{linear}
Let $\tilde{L}$ be a matrix given by $\tilde{L}=LS$, where $L$ is symmetric,
$S= diag \left( s_1, \cdots, s_N \right)$ nondegerate. Suppose that
all the eigenvalues
of $\tilde{L}$ be distinct. Then $\tilde{L}$ can be diagonalized
by a matrix $\Phi$ satisying (\ref{PhiS}), i.e.,
$\Phi S^{-1} \Phi^T = S^{-1}$,
$\Phi^T S \Phi = S$.
\end{Lemma}
\begin{Proof}
Let $\left(\cdot  ,\cdot \right)$ be the usual euclidean inner product, i.e.
$\left(x,y \right):=\sum_{i=1}^{N}x_iy_i$, for
$x=\left(x_1, \cdots, x_N \right)$ and
$y=\left(y_1, \cdots, y_N \right)$. Let $\phi(\lambda_i)$ be the eigenvector
of $\tilde{L}$ corresponding to
the eigenvalue $\lambda_i$, i.e., $\tilde{L}\phi(\lambda_i)=
\lambda_i\phi(\lambda_i)$. Calculating $(S\phi(\lambda_j),
\tilde{L}\phi(\lambda_i))$, we have
$$\ (S\phi(\lambda_j), \tilde{L}\phi(\lambda_i))=\lambda_i(S\phi(\lambda_j),
\phi(\lambda_i))$$
$$=(LS\phi(\lambda_j), S\phi(\lambda_i))
=\lambda_j(S\phi(\lambda_j), \phi(\lambda_i)),$$
which leads to
$$(\lambda_j-\lambda_i)(S\phi(\lambda_j), \phi(\lambda_i))=0.$$
By the assumption, $\lambda_i \neq \lambda_j$ for $i \neq j$, we obtain
$(S\phi(\lambda_j), \phi(\lambda_i))=0$, that is,
$\Phi^T S \Phi$ is a diagonal matrix. It is also nondegenerate.
We then normalize $\Phi$ to satisfy $\Phi^T S \Phi=S$.
We multiply $S^{-1}\Phi^T$ on the both
sides from the right, get $\Phi^TS\Phi S^{-1}\Phi^T=\Phi^T$. This leads to
$\Phi S^{-1} \Phi^T = S^{-1}$.
\end{Proof}

\medskip

{\bf Remark 2} With the normalization, the eigenmatrix $\Phi$ becomes complex
in general, even in the case that all the eigenvalues are real. This is simply
due to a case where the sign of $\sum_{k=1}^Ns_k\phi_k^2(\lambda_i)$
in $\Phi^TS\Phi$ differs
from that of $s_i$.

\medskip

We now show that the choice of $\tilde B_n$ in (\ref{comp2}) is compatible with
(\ref{PhiS}), that is:
\begin{Lemma}
Eqs
(\ref{PhiS}) are invariant under the flows generated
by $\tilde B_n=\prod_a \tilde L^n$.
\end{Lemma}
\begin{Proof}
First, we have
\begin{eqnarray*}
\frac{\partial (\Phi^T S \Phi)}{\partial t_n}
&=& \frac{\partial \Phi^T }{\partial t_n}S \Phi +
\Phi^T S\frac{\partial \Phi}{\partial t_n} \\
&=& \Phi^T(\tilde B_n^TS + S\tilde B_n)\Phi \ .
\end{eqnarray*}
So all we need to show is $\tilde B_n^TS + S\tilde B_n=0$.
Since $S$ is diagonal, it
commutes with the anti-symmetric projection $\prod_{a}$, that is,
\begin{eqnarray*}
\tilde B_n := (\tilde {L}^n)_{>0}-(\tilde {L}^n)_{<0} & = &
 (\tilde {L}^{n-1}LS)_{>0}- (\tilde {L}^{n-1}LS)_{<0} \\
& = & (\tilde {L}^{n-1}L)_{>0}S -(\tilde {L}^{n-1}L)_{<0}S.
\end{eqnarray*}
This leads to
\begin{eqnarray*}
\tilde B_n^TS + S\tilde B_n = S[(\tilde {L}^{n-1}L)_{>0} -
(\tilde {L}^{n-1}L)_{<0}]^TS \\
+ S[(\tilde {L}^{n-1}L)_{>0} - (\tilde {L}^{n-1}L)_{<0}]S.
\end{eqnarray*}
 Since
$\tilde {L}^{n-1}L$ is symmetric,
$(\tilde {L}^{n-1}L)_{>0} - (\tilde {L}^{n-1}L)_{<0}$ is skew-symmetric.
Thus, we have $\tilde B_n^TS + S\tilde B_n=0$.
\end{Proof}

\medskip

The eigenmatrix $\Phi$ consists of the eigenvectors of $\tilde L$,
say $\phi(\lambda_{k}) \equiv ( \phi_{1}(\lambda_{k}),\\\cdots $,
$\phi_{N}(\lambda_{k} ))^{T} $ for $k = 1, 2, \cdots , N$,
\begin{eqnarray}
\label{orth}
\Phi \equiv \left[ \phi(\lambda_{1}), \ \cdots \ , \ \phi(\lambda_{N})
\right] \ = \ \left[ \phi_{i}(\lambda_{j}) \right]_{1 \le i , j \le N} \ .
\end{eqnarray}
Then (\ref{PhiS}) give the ``orthogonality'' relations
\begin{eqnarray}
\label{ortho}
\sum_{k=1}^{N} s_k^{-1}\phi_{i}(\lambda_{k}) \phi_{j}(\lambda_{k}) =
\delta_{ij} s_i^{-1} \ , \\
\sum_{k=1}^{N} s_k\phi_{k}(\lambda_{i}) \phi_{k}(\lambda_{j}) =
\delta_{ij}s_i \ .
\end{eqnarray}
With (\ref{ortho}), we now define an inner product $<\cdot,\cdot>$ for
two function $f$ and $g$ of $\lambda$ as
\begin{eqnarray}
\label{product}
<f, g> := \sum_{k=1}^Ns_k^{-1}f(\lambda_k)g(\lambda_k),
\end{eqnarray}
which we write as $<fg>$ in the sequel.
The metric in the inner product is given by
\begin{eqnarray}
\label{metric}
d \alpha (\lambda) = \sum_{k=1}^{N} s_k^{-1}\delta(\lambda - \lambda_{k}) d
\lambda \ ,
\end{eqnarray}
which leads to an indefinite metric due to a choice of negative entries
$s_k$ in $S$.
The entries of $\tilde L^n$ are then expressed by
\begin{eqnarray}
\label{back}
\tilde a_{ij}^{(n)} := \left( \tilde L^{n} \right)_{ij} =
 s_j < \lambda^{n} \phi_{i} \phi_{j} > \ .
\end{eqnarray}

\section{Inverse scattering method}
\renewcommand{\theequation}{3.\arabic{equation}}\setcounter{equation}{0}

In this section, we solve the initial value problem of the hierarchy
(\ref{todals}) by using the inverse scattering method.
Namely, we first solve the time evolution for the eigenmatrix $\Phi({\bf t})$
of $\tilde L$ with the initial matrix
$\Phi(0) := \Phi^0=[\phi_i^0(\lambda_j)]_{1 \le i, j \le N}$ obtained
from the eigenvalue problem $\tilde L(0)\Phi^0=\Phi^0\Lambda$, and then
find the solution $\tilde L({\bf t})$ through (\ref{back}) with
$\Phi({\bf t})$.
Here we call the eigenvalues $\lambda_k$, $k=1,\cdots,N$ and the eigenmatrix
 $\Phi$ the ``scattering data''.

\medskip

Let us first note
$\tilde B_n=\tilde L^n-diag(\tilde L^n)-2(\tilde L^n)_{<0}$, and using
$\tilde L\Phi=\Phi \Lambda$, we write (\ref{comp2}) as
\begin{eqnarray}
\label{alterphi}
\frac{\partial}{\partial t_n}\Phi=\Phi\Lambda^n- \left[ diag(\tilde L^n)+
2(\tilde L^n)_{<0} \right]\Phi \ .
\end{eqnarray}
Then the
equations for the first row vector $\phi_1(\lambda_k,{\bf t})$,
$k=1, \cdots, N$,
in $\Phi({\bf t})$ are
\begin{eqnarray}
\label{phi1}
\frac{\partial \phi_1(\lambda_k)}{\partial t_n}=
\left( \lambda_k^n-s_1<\lambda^n\phi_1^2(\lambda)> \right) \phi_1(\lambda_k) \
,
\end{eqnarray}
which can be readily solved in the form
$\phi_1(\lambda_k,{\bf t})=\frac{\psi_1(\lambda_k,{\bf t})}
{\sqrt{s_1<\psi_1^2(\lambda, {\bf t})>}}$  with
\begin{eqnarray}
\label{psi1}
\psi_1(\lambda_k,{\bf t})= \phi_1^0(\lambda_k)e^{\xi(\lambda_k,{\bf t})},
\end{eqnarray}
where
$e^{\xi(\lambda_k,{\bf t})}:=\sum_{n=1}^\infty \lambda_k^nt_n$.
The equations for the second row vector $\phi_2(\lambda_k,{\bf t})$,
$k=1, \cdots, N$, are
\begin{eqnarray}
\label{phi2}
\nonumber
\frac{\partial \phi_2(\lambda_k)}{\partial t_n}
=\left(\lambda_k^n-s_2<\lambda^n\phi_2^2(\lambda)>\right) \phi_2(\lambda_k)\\
- 2s_1<\lambda^n\phi_2(\lambda)\phi_1(\lambda)>\phi_1(\lambda_k) \ .
\end{eqnarray}
The solution to (\ref{phi2}) can be also found in the form
$\phi_2(\lambda_k,{\bf t} )=\frac{\psi_2(\lambda_k,{\bf t})}
{\sqrt{s_2<\psi_2^2(\lambda, {\bf t})>}}$ with
\begin{eqnarray}
\label{psi2}
\psi_2(\lambda_k,{\bf t})=
\phi_2^0(\lambda_k)e^{\xi(\lambda_k,{\bf t})}-
s_1<\phi_2^0(\lambda)e^{\xi(\lambda,{\bf t})}\phi_1(\lambda,{\bf t})>
\phi_1(\lambda_k,{\bf t}).
\end{eqnarray}
In general, the $i$th row vector of $\Phi$ in (\ref{alterphi}) satisfies
\begin{eqnarray}
\label{phii}
\nonumber
\frac{\partial \phi_i(\lambda_k)}{\partial t_n}
=\left(\lambda_k^n-s_i<\lambda^n\phi_i^2(\lambda)> \right) \phi_i(\lambda_k)\\
-2\sum_{l=1}^{i-1}s_l<\lambda^n\phi_i(\lambda)
\phi_l(\lambda)>\phi_l(\lambda_k).
\end{eqnarray}
Then the solution of (\ref{phii}) is expected to have
the form
$\phi_i(\lambda_k,{\bf t})=\frac{\psi_i(\lambda_k,{\bf t})}
{\sqrt{s_i<\psi_i^2(\lambda, {\bf t})>}}$ with
\begin{eqnarray}
\label{psii}
\psi_i(\lambda_k, {\bf t})=
\phi_i^0(\lambda_k)e^{\xi(\lambda_k,{\bf t})}-
\sum_{l=1}^{i-1}s_l<\phi_i^0(\lambda)
e^{\xi(\lambda,{\bf t})}\phi_l(\lambda,{\bf t})>\phi_l(\lambda_k,{\bf t}) \ .
\end{eqnarray}
As we will see below, this is indeed the case.
The above procedure suggests that
the eigenmatrix
$\Phi$ can be constructed by the ``orthonormalization procedure'' on
the row vectors $[\phi_i^0(\lambda_k)
e^{\xi(\lambda_k,{\bf t})}]_{1 \le k \le N}, i=1, \cdots, N$.

\medskip

Let us now give a more systematic presentation of the above procedure, which
is the essence of the approach in \cite{KM}. Because of the simple
structure of $\psi_i(\lambda_k)$ in (\ref{psii}), we first write
\begin{eqnarray}
\label{trans}
\Phi = T \Psi \ ,
\end{eqnarray}
where $\Psi := [\psi_i(\lambda_j)]_{1 \le i, j \le N}$ and
\begin{eqnarray}
\nonumber
T = diag \left[ (s_1< \psi_{1}^{2}>)^{-1/2}\ , \cdots \ , \
(s_N< \psi_{N}^{2}>)^{-1/2} \right] \ .
\end{eqnarray}
Note that (\ref{trans}) can be regarded as a gauge transformation and
includes a freedom in the
choice of $\psi$. Namely, (\ref{trans}) is invariant under
the scaling of $\psi_{i}$, i.e., $\psi_{i}
\rightarrow f_{i}({\bf t}) \psi_{i}$, with $\{f_{i}\}_{i=1}^{N}$ arbitrary
functions of ${\bf t}$.  With (\ref{trans}), (\ref{comp1}) and (\ref{comp2})
become
\begin{eqnarray}
\label{lam1}
\left(
T^{-1} \tilde L T \right) \Psi &=& \Psi \Lambda \ , \\
\label{lam2}
\frac{ \partial}{\partial t_{n}}\Psi &=&
\left( T^{-1} \tilde B_{n} T \right) \Psi -
\left(\frac{ \partial}{\partial t_{n}} \log{T} \right)\Psi \ .
\end{eqnarray}
Writing as in (\ref{alterphi}), i.e.
\begin{eqnarray}
\nonumber
\left( T^{-1} \tilde B_{n} T \right) =
 -2 \left( T^{-1} \tilde L^{n} T \right)_{<0} +
\left( T^{-1} \tilde L^{n} T \right) - diag\left( \tilde L^{n} \right) \ ,
\end{eqnarray}
(\ref{lam2}) gives
\begin{eqnarray}
\label{nlam}
\frac{ \partial \psi}{\partial t_{n}} =
 -2 \left( T^{-1} \tilde L^{n} T \right)_{<0} \psi +
\lambda^{n} \psi - \left( diag\left( \tilde L^{n} \right) +
\frac{ \partial }{\partial t_{n}} \log{ T } \right) \psi \ .
\end{eqnarray}
We here observe that (\ref{nlam}) can be split into two sets of
equations by fixing the guage freedom in the determination of $\psi$.  In
the components, these are
\begin{eqnarray}
\label{cpn1}
\frac{ \partial \psi_i}{\partial t_{n}} =
- 2 \sum_{j=1}^{i-1} \frac{ < \lambda^{n} \psi_{i} \psi_{j} >}
{< \psi_{j}^{2} >} \psi_{j} + \lambda^{n} \psi_{i} \ ,
\end{eqnarray}
\begin{eqnarray}
\label{cpn2}
\frac{1}{2} \frac{ \partial }{\partial t_{n}} \log{
< \psi_{i}^{2} > } = \tilde a_{ii}^{(n)} \ .
\end{eqnarray}
It is easy to check that (\ref{cpn1}) implies (\ref{cpn2}).
It is also immediate from (\ref{cpn1}) that we have:
\begin{Proposition}
The solution of (\ref{cpn1}) can be written in the form of separation
of variables,
\begin{eqnarray}
\label{sepa}
\psi(\lambda,{\bf t}) = A({\bf t}) \phi^{0}(\lambda)
e^{\xi(\lambda,{\bf t})} \ ,
\end{eqnarray}
where $A({\bf t})$ is a lower triangular matrix with $diag[A({\bf t})]
= A({\bf t}=0) = I_{N}$, the $N \times N$ identity matrix, and
$\phi^{0}({\lambda}) = \phi(\lambda, 0)$.
\end{Proposition}
Note that we have chosen the initial conditions of $\psi(\lambda,{\bf t})$
to coincide with $\phi(\lambda, {\bf t})$, i.e.  $\psi(\lambda,0) =
\phi^{0}(\lambda)$, and thus
$s_i<\psi_{i} \psi_{j}>({\bf t}=0) = s_i<\phi_{i}^{0}
\phi_{j}^{0} > =
\delta_{ij}$.  As a direct consequence of this proposition, and the
orthogonality of the eigenvectors, (\ref{ortho}), i.e. $< \psi_{i}
\psi_{j} > =0$ for $i \neq j$, we have:
\begin{Corollary}(Orthogonality):
For each $i \in \{2, \cdots, N\}$, we have for all ${\bf t}$ with
$t_{m} \in {\bf R}$,
\begin{eqnarray}
\label{corr}
<\psi_{i} \phi_{j}^{0} e^{\xi(\lambda,{\bf t})}> \equiv \sum_{k=1}^{N}
s_k^{-1}\psi_{i}(\lambda_{k},{\bf t}) \phi_{j}^{0}(\lambda_{k})
e^{\xi(\lambda_{k},{\bf t})} = 0
\end{eqnarray}
for $j = 1, 2, \cdots, i-1$.
\end{Corollary}

Now we obtain the formula
for the eigenvectors of $\tilde L$ in terms of the initial data $\{
\phi_{i}^{0}(\lambda) \}_{1 \le i
\le N}$:
\begin{Theorem}
The solutions  $\psi_{i}(\lambda, {\bf t})$ of (\ref{cpn1}) are given by
\begin{eqnarray}
\label{psis}
\psi_{i}(\lambda,{\bf t}) = \frac{e^{\xi(\lambda,{\bf t})}}
{ D_{i-1}({\bf t})} \left|
\begin{array}{ccc}
s_1c_{11} & \ldots & s_1c_{1i} \\
\vdots & \ddots & \vdots \\
s_{i-1}c_{i-1 1} & \ldots & s_{i-1}c_{i-1 i} \\
\phi_{1}^{0}(\lambda) & \ldots & \phi_{i}^{0}(\lambda) \\
\end{array}
\right|
\end{eqnarray}
where $c_{ij}({\bf t}) = < \phi_{i}^{0} \phi_{j}^{0} e^{ 2
\xi(\lambda,{\bf t})}>$, and $D_{k}({\bf t})$ is the determinant of the
$k \times k$ matrix with entries $s_ic_{ij}({\bf t})$, i.e.,
\begin{eqnarray}
\label{DDD}
D_{k}({\bf t}) = \left| \Big( s_ic_{ij}({\bf t}) \Big)_{1 \le i,j \le k}
\right| \ .
\end{eqnarray}
(Note here that $s_ic_{ij}(0)=\delta_{ij}$ and $D_k(0)=1$.)
\end{Theorem}
\begin{Proof}
{}From equation (\ref{corr}) with (\ref{sepa}), we have
\begin{eqnarray}
\label{pro}
s_l\sum_{ k = 1}^{i} A_{ik}({\bf t}) < \phi_{k}^{0} \phi_{\ell}^{0} e^{2
\xi(\lambda,{\bf t})} > = 0 \ , \ for \ 1 \le \ell \le i-1 \ .
\end{eqnarray}
Solving (\ref{pro}) for $A_{ik}$ with $A_{ii} = 1$, we obtain
\begin{eqnarray}
A_{ik}({\bf t}) = - \frac{ D_{i-1}^{k}({\bf t})}{D_{i-1}({\bf t})} \ ,
\end{eqnarray}
where $D_0({\bf t}):=1$, and $D_{i-1}^{k}({\bf t})$ is the determinant
$D_{i-1}({\bf t})$ in the form
(\ref{DDD}) with the replacement of the kth column $\left( s_1c_{1k}, \ldots  ,
 s_{i-1}c_{i-1 \ k} \right)^{T}$ by $\left( s_1c_{1i}, \ldots  ,
s_{i-1}c_{i-1 \ i} \right)^{T}$.  From (\ref{sepa}), we then have
\begin{eqnarray}
\nonumber
\psi_{i} &=& e^{\xi(\lambda,{\bf t})} \sum_{k=1}^{i} A_{ik}\phi_{k}^{0} \\
\nonumber
&=& - e^{\xi(\lambda,{\bf t})}\sum_{k=1}^{i-1} \frac{D_{i-1}^{k}
\phi_{k}^{0}}{D_{i-1}} + e^{\xi(\lambda,{\bf t})}
\frac{D_{i-1} \phi_{i}^{0}}{D_{i-1}} \\
\nonumber
&=& \frac{e^{\xi(\lambda,{\bf t})}}{D_{i-1}} \times \left\{
 - \phi_{1}^{0}
\left|
\begin{array}{cccc}
s_1c_{1i} & s_1c_{12} & \cdots & s_1c_{1 i-1} \\
\vdots & \ddots & \ddots & \vdots \\
s_{i-1}c_{i-1 i} & s_{i-1}c_{i-1 2} & \cdots & s_{i-1}c_{i-1 i-1} \\
\end{array} \right|
+ \cdots \right.  \\
\nonumber
& &  \hspace{.3in} \left. \cdots +
 \phi_{i}^{0}
\left|
\begin{array}{ccc}
s_1c_{11} & \cdots & s_1c_{1 i-1} \\
\vdots & \ddots & \vdots \\
s_{i-1}c_{i-1 1} & \cdots & s_{i-1}c_{i-1 i-1} \\
\end{array} \right| \right\} \ ,
\end{eqnarray}
which is just (\ref{psis}).
\end{Proof}

\medskip

We note from (\ref{psis}) that $<\psi_{i}^{2}>$ can be expressed by
$D_{i}$,
\begin{eqnarray}
< \psi_{i}^{2} > = \frac{D_{i}}{s_iD_{i-1}} \ .
\end{eqnarray}
This yields the formulae for the normalized eigenfunctions
\begin{eqnarray}
\label{evcs}
\phi_{i}(\lambda,{\bf t}) = \frac{e^{\xi(\lambda,{\bf t})}}
{\sqrt{D_{i}({\bf t})
D_{i-1}({\bf t})}} \left|
\begin{array}{ccc}
s_1c_{11} & \ldots & s_1c_{1i} \\
\vdots & \ddots & \vdots \\
s_{i-1}c_{i-1 1} & \ldots & s_{i-1}c_{i-1 i} \\
\phi_{1}^{0}(\lambda) & \ldots & \phi_{i}^{0}(\lambda) \\
\end{array}
\right|
\end{eqnarray}
With the formula (\ref{evcs}), we now have the solution (\ref{back})
of the inverse scattering problem (\ref{comp1}) and (\ref{comp2}).

\medskip

The above derivation of (\ref{evcs}) is the same as the orthogonalization
procedure of Szeg{\"{o}} \cite{Sz39},
which is equivalent to the Gram - Schmidt
orthogonalization as observed in \cite{KM}, except here the orthogonalization
is with respect to an indefinite metric.
Indeed, from the form of $\phi_{i} =
\frac{\psi_{i}} {\sqrt{s_i< \psi_{i}^{2} >}}$, we see that
(\ref{sepa}) expresses the relations
\begin{eqnarray}
\label{spa}
\phi_{i}(\lambda,{\bf t}) \in span \left\{
\phi_{1}^{0}(\lambda)e^{\xi(\lambda,{\bf t})}, \cdots,
\phi_{i}^{0}(\lambda)e^{\xi(\lambda,{\bf t})} \right\} \ , \ i=1, \cdots, N,
\end{eqnarray}
where we are viewing $\{ \phi_{i}(\lambda,{\bf t}) \}_{i=1}^{N}$ as a
collection of ``orthogonal'' functions defined on the eigenvalues of the
matrix $\tilde L$.
The relations (\ref{spa}) together with the orthogonality (\ref{ortho}),
$<\phi_{i} \phi_{j}> = s_i^{-1}\delta_{ij} $,
imply that $\left\{ \phi_{i}(\lambda,{\bf t})
\right\}_{i=1}^{N}$ are obtained by an orthogonalization of the sequence
$\left\{ \phi_{i}^{0}(\lambda)e^{\xi(\lambda,{\bf t})} \right\}_{i=1}^{N}$,
and hence one may obtain (\ref{evcs}) from the classical
formulae as in \cite{Sz39}.
If in the orthogonalization procedure for $\phi_{i}(\lambda)$, we
replace $\left\{ \phi_{1}^{0}(\lambda)e^{\xi(\lambda,{\bf t})}, \cdots,
\phi_{i-1}^{0}(\lambda)e^{\xi(\lambda,{\bf t})} \right\}$ by
$\left\{\phi_{1}(\lambda, {\bf t}), \cdots,
\phi_{i-1}(\lambda, {\bf t}) \right\}$,
then we get the solutions given by (\ref{psii}).

\medskip

{\bf Remark 3}.  The above method is a natural generalization of the QR
factorization method of Symes \cite{symes}. In the case of $L$ being symmetric,
the QR factorization method is given as follows: First factorize $e^{L(0)t}=
Q(t)R(t)$, where $Q(t)\in SO(N)$ with $Q(0)=I_N$, and $R(t)$ is a lower
triangular matrix. Then $L(t)$ is given by $L(t)=Q^{-1}(t)L(0)Q(t)$.
In the case of our $\tilde L$, the same procedure applies where $Q(t)$
satisfies
(\ref{PhiS}), i.e. $\Phi(t) = Q^{-1}(t) \Phi^0$.

\medskip

{\bf Remark 4}. The above method applies directly to the following equation,
\begin{eqnarray*}
\frac{\partial }{\partial t_{1}} \tilde L = \ \left[ \tilde B_{1} \ ,
\ \tilde L \right]\ + \ f(\tilde L),
\end{eqnarray*}
where a ``pumping'' term $f(\tilde L)$ is given by
an entire function of $\tilde L$.
Due to the presence of $f(\tilde L)$,
the eigenvalues of $\tilde L(t)$
are no longer time-independent, In fact,
we have $$\lambda_{t_1}=f(\lambda),$$ and
the solution formula (\ref{evcs}) still holds
with the replacement of $e^{\lambda t_1}$ by
$\exp \left(\int_0^{t_1} {\lambda (t)dt} \right)$,
and $t_i=0$ except $i=1$.

\section{Behaviors of the solutions}
\renewcommand{\theequation}{4.\arabic{equation}}\setcounter{equation}{0}

Here we consider the first equation of the hierarchy, that is, we have only
one time $t_1$, which we denote by $t$. As a consequence of the explicit
formula (\ref{evcs}), we can now study the behaviors of the solutions.
First we note:
\begin{Proposition}
\label{real}
$D_i, i = 1, 2, \cdots, N$, in (\ref{DDD}) are real functions.
\end{Proposition}
\begin{Proof}
In the construction of the solutions $\Phi(t)$,
the``gauge'' $T$ is fixed by (\ref{cpn2}).
In terms of $D_i$, (\ref{cpn2}) is
\begin{eqnarray}
\label{cpn3}
\tilde a_{ii}=\frac{1}{2} \frac{ \partial }{\partial t} \log{
\frac{D_i}{D_{i-1}}}  .
\end{eqnarray}
Note that
$D_0 \equiv 1$, $D_i(0)=1$ and $ \tilde a_{ii}$ are real functions. Then
we see by
induction that all $D_i$ are real functions.
\end{Proof}

\medskip

In (\ref{cpn3}), suppose $D_i(t_0)=0$ for some finite $t_0$ and some $i$.
Then if $ \tilde L(t_0)$ is a finite matrix, $D_{i-1}(t_0)$ must be also 0.
By induction, $D_1(t_0)=0$, but $D_0(t) \equiv 1 $, this forces
$\tilde a_{11}$ to be infinite, which is a contradiction. So we have:
\begin{Proposition}
\label{zero}
Suppose $D_i(t_0)=0$ for some $t_0 < \infty$ and some $i$, then
$ \tilde L(t)$ blows up to infinity at $t_0$.
\end{Proposition}

We note that
$D_i, i = 1, 2, \cdots, N$, are $i$th principal minors of the
product of matrices $S \Phi_e S^{-1} \Phi_e^T$, where
$\Phi_e$ is
\begin{eqnarray*}
\left(
\begin{array}{ccc}
e^{\lambda_1 t}\phi_{1}^{0}(\lambda_1) & \ldots &
e^{\lambda_N t}\phi_{1}^{0}(\lambda_N) \\
\vdots & \ddots & \vdots \\
e^{\lambda_1 t}\phi_{N}^{0}(\lambda_1) & \ldots &
e^{\lambda_N t}\phi_{N}^{0}(\lambda_N)
\end{array}
\right) .
\end{eqnarray*}
Then from the Cauchy-Binet theorem, we have:
\begin{Proposition}
\label{expansion}
$D_i, i = 1, 2, \cdots, N$ can be expressed as
\begin{eqnarray}
\label{expand}
D_i = s_1\cdots s_i\sum_{J_N=(j_1,\cdots,j_i)_N}
\frac{1}{s_{j_1}\cdots s_{j_i}}
e^{2\sum_{k=1}^i\lambda_{{j_k}}t} \left|
\begin{array}{ccc}
\phi_1^0(\lambda_{j_1}) & \ldots & \phi_1^0(\lambda_{j_i}) \\
\vdots & \ddots & \vdots \\
\phi_i^0(\lambda_{j_1}) & \ldots & \phi_i^0(\lambda_{j_i}) \\
\end{array}
\right|^2 ,
\end{eqnarray}
where $J_N$ represents all possible combinations for
$1\le j_1<\cdots<j_i\le N$.
In particular $D_0(t) \equiv 1$, and $D_N(t)=\exp(2 \sum_{i=1}^N\lambda_it)$.
\end{Proposition}
This proposition is very useful to study the asympototic behavior of
$D_i$ for large $t$.
To study the time evolution of $D_i(t)$, we need the information on the
scattering data, that is, the eigenvalues $\lambda_k$ and the eigenmatrix
$\Phi^0$. Here we have from the linear algebra:

\begin{Lemma}
\label{ls}
Let $\tilde{L}$ be a matrix given by $\tilde{L}=LS$, where $L$ is symmetric,
$S= diag \left( s_1, \cdots, s_N \right)$ positive definite. Then all the
eigenvalues of $\tilde{L}$ are real, and the initial eigenmatrix
$\Phi^0$ of $\tilde{L}$ satisfying (\ref{PhiS}) is also real.
\end{Lemma}
\begin{Proof}
Let $\phi (\lambda_i)$ be an eigenvector of $\tilde{L}$ corresponding to
$\lambda_i$, i.e.,
\begin{eqnarray}
\label{eigen}
\tilde{L}\phi (\lambda_i) = \lambda_i \phi (\lambda_i).
\end{eqnarray}
Multiplying (\ref{eigen}) by the adjoint of $S\phi(\lambda_i)$, i.e.,
$\bar \phi^T (\lambda_i)S$ to the left, we get
\begin{eqnarray}
\label{eigen1}
\bar \phi^T (\lambda_i)SLS\phi (\lambda_i) = \lambda_i
\bar \phi^T (\lambda_i) S \phi (\lambda_i).
\end{eqnarray}
On the other hand, if we take the adjoint of (\ref{eigen}), and then
multiply $S\phi (\lambda_i)$ to the right, we get
\begin{eqnarray}
\label{eigen2}
\bar \phi^T (\lambda_i)SLS\phi (\lambda_i) = \bar \lambda_i
\bar \phi^T (\lambda_i) S \phi (\lambda_i).
\end{eqnarray}
Subtracting (\ref{eigen1}) from (\ref{eigen2}), we have
$(\lambda_i-\bar \lambda_i)\bar \phi^T (\lambda_i) S \phi (\lambda_i)=0$.
Since S is positive definite, this forces $\lambda_i-\bar \lambda_i=0$,
that is, $\lambda_i$ is real.

As for the normalization of
$\phi (\lambda_i)$,
since both $\phi^T (\lambda_i) S \phi (\lambda_i)$ and $s_i$
are positive, we only need to multiply $\phi (\lambda_i)$ by
some ``real'' factor. So $\phi (\lambda_i)$ remains real after normalization.
This verifies the assertion of the lemma.
\end{Proof}

\medskip

We now obtain:
\begin{Theorem}
Let the eigenvalues of $\tilde L$ be ordered as $\lambda_1>\lambda_2>\cdots>
\lambda_N$.
Suppose that $S$ is positive definite and $det \Phi_{n}^{0} \neq 0$ for $ n =
1, \ldots , N$, where $\Phi_{n}^{0}$ is the $n$th principal minor of
$\Phi^{0}$.  Then
as $t \rightarrow \infty$, the eigenfunctions
$\phi_{i}(\lambda_{j},t)$ satisfy
\begin{eqnarray}
\label{dell1}
\phi_{i}(\lambda_{j},t) \rightarrow \delta_{ij} \times
sgn \left( det \Phi_{i}^{0}\right) sgn\left( det \Phi_{i-1}^{0} \right)  \ ,
\end{eqnarray}
which implies $L(t) \rightarrow diag \left(\lambda_{1}, \ldots ,
\lambda_{N} \right)$ from (\ref{comp1}).
\end{Theorem}
\begin{Proof}
Using Lemma 3, we see that all the terms in the sum (\ref{expand}) are
positive, thereby $D_i$'s are positive for all $t$.
{}From the ordering in the eigenvalues, we see that
the leading order for $D_i$ is given by
\begin{eqnarray}
\label{leading}
e^{2 \sum_{k=1}^i\lambda_{{k}}t} \left|
\begin{array}{ccc}
\phi_1^0(\lambda_{1}) & \ldots & \phi_1^0(\lambda_{i}) \\
\vdots & \ddots & \vdots \\
\phi_i^0(\lambda_{1}) & \ldots & \phi_i^0(\lambda_{i}) \\
\end{array}
\right|^2 ,
\end{eqnarray}
where we have assumed $det\Phi_i^0 \neq 0$. From (\ref{evcs})
and (\ref{leading}), as $t \rightarrow \infty$,
\begin{eqnarray}
\nonumber
& &\phi_{n}(\lambda; t) \rightarrow \frac{ e^{\lambda t}}
{\left| det \Phi_{n}^{0} \right| \left| det \Phi_{n-1}^{0} \right| \exp{
\left[ \left( 2 \sum_{i=1}^{n-1} \lambda_{i} + \lambda_{n} \right) t
\right]}} \times \\
\label{blt}
& & \hspace{.6in} \times  \left|
\begin{array}{ccc}
s_1c_{11}(t) & \cdots & s_1c_{1n}(t) \\
\vdots & \ddots & \vdots \\
s_{n-1}c_{1 n-1}(t) & \cdots & s_{n-1}c_{n-1 n} \\
\phi_{1}^{0}(\lambda) & \cdots & \phi_{n}^{0}(\lambda) \\
\end{array} \right| \ .
\end{eqnarray}
The dominant term in the determinant gives
\begin{eqnarray}
\nonumber
& &
e^{2 \sum_{i=1}^{n-1} \lambda_{i}t} \sum_{{\bf P}_{n-1}}
\phi_{1}^{0}(\lambda_{\ell_{1}}) \cdots
\phi_{n-1}^{0}(\lambda_{\ell_{n-1}})
\left|
\begin{array}{ccc}
\phi_{1}^{0}(\lambda_{\ell_{1}}) & \ldots &
\phi_{n}^{0}(\lambda_{\ell_{1}}) \\
\vdots & \ddots & \vdots \\
\phi_{1}^{0}(\lambda_{\ell_{n-1}}) & \ldots &
\phi_{n}^{0}(\lambda_{\ell_{n-1}}) \\
\phi_{1}^{0}(\lambda) & \ldots &
\phi_{n}^{0}(\lambda) \\
\end{array} \right| \\
\label{thdo}
& & =
e^{2 \sum_{i=1}^{n-1} \lambda_{i}t}
\left|
\begin{array}{ccc}
\phi_{1}^{0}(\lambda_{1}) & \ldots &
\phi_{n}^{0}(\lambda_{1}) \\
\vdots & \ddots & \vdots \\
\phi_{1}^{0}(\lambda_{n-1}) & \ldots &
\phi_{n}^{0}(\lambda_{n-1}) \\
\phi_{1}^{0}(\lambda) & \ldots &
\phi_{n}^{0}(\lambda) \\
\end{array} \right| \times \\
\nonumber
& & \hspace{1.0in} \times
\sum_{{\bf P}_{n-1}}
\phi_{1}^{0}(\lambda_{\ell_{1}}) \cdots
\phi_{n-1}^{0}(\lambda_{\ell_{n-1}}) \sigma(\{ \ell_{j} \}_{j=1}^{n-1}) \ ,
\end{eqnarray}
where $\sigma(\{ \ell_{j} \}_{j=1}^{n-1})$ is the signature of the
permutation ${\bf P}_{n-1}$ for $1 \le l_j \le n-1$.  We note that the
determinant in (\ref{thdo}) is
zero for $\lambda = \lambda_{j}$, $j=1, \ldots, n-1$.  The ordering
$\lambda_{1} > \cdots > \lambda_{N}$ implies the result.
\end{Proof}

\medskip

Theorem 2 shows if $S$ is positive definite, then generic
solutions have the ``sorting property''.
It is a natural generalization of Theorem 2 in \cite{KM},
where $S=I_N$. Next theorem provides sufficient
conditions for the solutions to blow up to infinity in finite time.
\begin{Theorem}
Suppose some eigenvalues of $\tilde{L}$ are not real, or $\Phi^0$ is not
real, and $det\Phi_n^0\neq 0$ for $n=1, \cdots, N$. Then $\tilde{L}(t)$
blows up to infinity in finite time.
\end{Theorem}
\begin{Proof}
We have two cases to consider:

\medskip

a). All the eigenvalues are real, but $\Phi^0$ is complex.\\
{}From Remark 1, each column $\phi^0 (\lambda_i)$ of $\Phi^0$ is either pure
imaginary or real. In this case, let $k$ be the first column such that
$\phi^0 (\lambda_k)$ is pure imaginary. With
the ordering $\lambda_{1} > \cdots > \lambda_{N}$, the leading order
in the expansion (\ref{expand}) of $D_k$ is still given by (\ref{leading}).
We also
assume $det\Phi_k^0 \neq 0$. Since $\phi^0 (\lambda_k)$ is pure imaginary,
while $\phi^0 (\lambda_i), i=1, \cdots, k-1$, are real, we have
\begin{eqnarray*}
(det\Phi_k^0)^2= \left|
\begin{array}{ccc}
\phi_1^0(\lambda_{1}) & \ldots & \phi_1^0(\lambda_{k}) \\
\vdots & \ddots & \vdots \\
\phi_k^0(\lambda_{1}) & \ldots & \phi_k^0(\lambda_{k}) \\
\end{array}
\right|^2 < 0
\end{eqnarray*}
Being negative of the leading order in $D_k$ implies $\lim_{t \rightarrow
\infty}
D_k(t)=-\infty$. Note $D_k(0)=1$, so there is some $t_0 < \infty$,
such that $D_k(t_0)=0$.
Then Proposition 4 results that $\tilde L(t)$ blows up to infinity in finite
time $t_0$.

\medskip

b). Some eigenvalues are not real.\\
We order the eigenvalues of $\tilde L$ by their real parts. We still
assume all the eigenvalues to be distinct. Since $\tilde L$ is a real
matrix, the complex eigenvalues appear as pairs. For a convenience,
we also assume that there is at most one pair having the same real part.
Suppose $\lambda_k+i\beta_k$ and $\lambda_k-i\beta_k$ are the first pair
of complex eigenvalues. Then from  (\ref{expand}),
the leading order term in $D_k$ is
\begin{eqnarray*}
\label{complex}
e^{2 \sum_{l=1}^k\lambda_{{l}}t+ 2i\beta_kt} \left|
\begin{array}{ccc}
\phi_1^0(\lambda_{1}) & \ldots & \phi_1^0(\lambda_{k}+i\beta_k) \\
\vdots & \ddots & \vdots \\
\phi_k^0(\lambda_{1}) & \ldots & \phi_k^0(\lambda_{k}+i\beta_k) \\
\end{array}
\right|^2 + \\
e^{2 \sum_{l=1}^k\lambda_{{l}}t- 2i\beta_kt} \left|
\begin{array}{ccc}
\phi_1^0(\lambda_{1}) & \ldots & \phi_1^0(\lambda_{k}-i\beta_k) \\
\vdots & \ddots & \vdots \\
\phi_k^0(\lambda_{1}) & \ldots & \phi_k^0(\lambda_{k}-i\beta_k) \\
\end{array}
\right|^2 .
\end{eqnarray*}
Since $D_k$ is real by Proposition 1, one can write the above as
$$e^{2 \sum_{l=1}^k\lambda_{l}t} \left[ A\cos(2 \beta_kt)+B\sin(2 \beta_kt)
\right].$$
where A and B are two real constants.
The above is an oscillating function about zero.
Thus by Proposition 2, we conclude that
$\tilde L(t)$ blows up to infinity in finite time.
\end{Proof}

\medskip

This theorem implies that the complexness of the initial
scattering data leads to the
blowing up of the solutions.
Now, only the situation left undetermined is the case where
$S$ is indefinite, and the scattering data are real. We then have:
\begin{Proposition}
\label{alternative}
Suppose conditions as stated above are satisfied and $det\Phi_i^0 \neq 0$,
$i=1,\cdots, N$.
If $\tilde L(t)$ doesn't blow up to infinity in finite time, then
$\tilde L(t)$ has the sorting property, i.e., $\tilde L(t) \rightarrow
diag(\lambda_1, \cdots, \lambda_N)$ as $t \rightarrow \infty$.
\end{Proposition}
\begin{Proof}
If $\tilde L(t)$ doesn't blow up to infinity in finite time, then
$D_i(t)$'s are all positive definite, in particular, the leading order in
the expansion (\ref{expand}) is positive. Then the same proof as in Theorem 3
applies here.
\end{Proof}

\section{Examples}
\renewcommand{\theequation}{5.\arabic{equation}}\setcounter{equation}{0}

We here demonstrate our results using two simple examples. The first
example includes a parameter, and we show a bifurcation behavior of the
solutions as the parameter changes. The second example is for a blowing
up aspect in the case where $S$ is indefinite, and all
eigenvalues of $\tilde L$ are real and so is $\Phi^0$.

\medskip

{\it Example 1}. We take $\tilde L$ to be a $2\times 2$
matrix with $S=diag(1, -1)$,
that is, $\Phi \in O(1,1)$,
and $\tilde L$ is given by
\begin{eqnarray}
\label{tildel}
\left(
\begin{array}{cc}
a_1 & -b_1 \\
b_1 & -a_2 \\
\end{array}
\right).
\end{eqnarray}
Our hierarchy (\ref{todals}) then gives for n=1
\begin{eqnarray}
\label{equation}
\left\{
\begin{array}{c}
\frac{da_1}{dt} = -2b_1^2, \\
\frac{da_2}{dt} = -2b_1^2, \\
\frac{db_1}{dt} = -b_1(a_1+a_2). \\
\end{array}
\right.
\end{eqnarray}
{}From (\ref{equation}), we can see both $a_1$ and $a_2$ are always decreasing.
If $a_1+a_2$ is initially negative, then $b_1^2$ increases
faster and faster, and we expect a blowing up of the system.
While, for the case where $a_1+a_2$ is
initially large positive,
we expect that $b_1^2$ to
decrease to 0 and $a_1+a_2$ decreases to some positive number. This is indeed
the case.

\medskip

Let us take the initial conditions, $a_1=0$, $b_1=1$, and $a_2=c$,
a constant.
We then show a bifurcation behavior of the solutions with the parameter
c, i.e., blowing up at $c<c_0$ and the sorting at $c>c_0$ for certain value
of $c_0$.
The initial scattering data, the eigenvalues $\lambda_{1,2}$ and the
normalized eigenmatrix $\Phi^0$, are given by
\begin{eqnarray}
\label{eigenvalues}
\lambda_{1,2}=\frac{1}{2} \left[ -c\pm\sqrt{c^2-4} \right],
\end{eqnarray}
and
\begin{eqnarray}
\label{phi0}
\Phi^0=\left(
\begin{array}{cc}
\frac{\lambda_2}{\sqrt{\lambda_2^2-1}}&
\frac{\lambda_1}{\sqrt{1-\lambda_1^2}} \\
\frac{-1}{\sqrt{\lambda_2^2-1}} & \frac{-1}{\sqrt{1-\lambda_1^2}} \\
\end{array}
\right).
\end{eqnarray}
The eigenvalues
$\lambda_1$ and $\lambda_2$ take the following values as the function of $c$:
$0>\lambda_1\ge-1\ge\lambda_2$, for $c\ge 2$; $\lambda_1=\bar \lambda_2$
complex for $|c|<2$; and $\lambda_1\ge1\ge\lambda_2>0$ for $c\le -2$.
Note in particular that $\Phi^0$ becomes complex (pure imaginary) when
$c<-2$, even though the eigenvalues are real.
Then from
(\ref{DDD}) and (\ref{evcs}), we have
\begin{eqnarray}
\label{d1}
D_1(t)=\frac{1}{\lambda_2-\lambda_1} \left( \lambda_2e^{2\lambda_1t}-\lambda_1
e^{2\lambda_2t} \right),
\end{eqnarray}
and
\begin{eqnarray}
\label{phit}
\Phi(t)=\frac{1}{\sqrt{D_1(t)}}
\left(
\begin{array}{cc}
\frac{\lambda_2}{\sqrt{\lambda_2^2-1}}e^{\lambda_1t} &
\frac{\lambda_1}{\sqrt{1-\lambda_1^2}}e^{\lambda_2t} \\
\frac{-1}{\sqrt{\lambda_2^2-1}}e^{\lambda_2t}
& \frac{-1}{\sqrt{1-\lambda_1^2}}e^{\lambda_1t} \\
\end{array}
\right).
\end{eqnarray}
With the chioce of $\lambda_1$ and $\lambda_2$ in (\ref{eigenvalues}),
we have $\lambda_1>\lambda_2$ for all $|c|>2$. In particular, notice
that the dominant term in $D(t)$ for large $t$ becomes negative for the
case $c<-2$. This is due to the complexness of $\Phi^0$, and implies
the blowing up in the solution at the time $t=t_B$ (Theorem 3),
\begin{eqnarray}
\label{tb}
t_B=\frac{1}{2(\lambda_1-\lambda_2)}\log \frac{\lambda_1}{\lambda_2}.
\end{eqnarray}
On the other hand, for $c>2$ we have the sorting result. For $|c|<2$,
the eigenvalues become complex and $D_1(t)$ is expressed as
\begin{eqnarray}
\label{d1t}
D_1(t)=-2(4-c^2)^{-\frac{1}{2}}e^{-2ct}\sin \left( \sqrt{4-c^2}t-\theta \right)
\end{eqnarray}
with $\tan \theta=\sqrt{4-c^2}/2$. This also indicates the blowing up
(Theorem 3). The solution $\tilde L(t)$ is obtained from (\ref{back}),
\begin{eqnarray}
\label{atilde}
\tilde a_{11}=a_1=<\lambda\phi_1^2>=\frac{e^{2\lambda_1t}-e^{2\lambda_2t}}
{\lambda_2e^{2\lambda_1t}-\lambda_1e^{2\lambda_2t}},
\end{eqnarray}
\begin{eqnarray}
\label{btilde}
\tilde a_{21}=b_1=<\lambda\phi_2\phi_1>
=\frac{(\lambda_2-\lambda_1)e^{(\lambda_1+\lambda_2)t}}
{\lambda_2e^{2\lambda_1t}-\lambda_1e^{2\lambda_2t}}.
\end{eqnarray}

\medskip

It is interesting to note that for the case $c\le -2$, the blowing up
occurs at one time $t_B$ (\ref{tb}), and the solution $\tilde L(t)$ will
be sorted as $t\rightarrow \infty$, with the asymptotic form,
\begin{eqnarray}
\label{asm}
\Phi (t) \rightarrow
\left(
\begin{array}{cc}
1 &
0 \\
0
& -1 \\
\end{array}
\right).
\end{eqnarray}
Also note that if we start with a ``wrong'' ordering in the eigenvalues,
i.e., $\lambda_1<\lambda_2$, then we still have the correct sorting result with
\begin{eqnarray}
\label{asmt}
\Phi (t) \rightarrow
\left(
\begin{array}{cc}
0 &
i \\
-i
& 0 \\
\end{array}
\right).
\end{eqnarray}
For the case of $c=\pm2$, we obtain by taking the limit of (\ref{atilde})
and (\ref{btilde})
\begin{eqnarray}
\label{ltilde}
\tilde L(t)=\left(
\begin{array}{cc}
-\frac{2t}{1\pm 2t} &
-\frac{1}{1\pm 2t} \\
\frac{1}{1\pm 2t}
& \mp2+\frac{2t}{1\pm 2t} \\
\end{array}
\right).
\end{eqnarray}
which showes the ``sorting property'' as $t\rightarrow \infty$, i.e.
$\tilde L(t) \rightarrow \pm diag(1,1)$. It should be noted that $\tilde L(0)$
at $c=\pm 2$ is not similar to $\pm diag(1,1)$.

\medskip

In the above example, $S$ is indefinite, when $c>2$, the eigenvalues are real
and
so is $\Phi^0$, the system has the sorting property. The following example
shows the blowing up aspect in that situation.

\medskip

{\it Example 2}. Take $S=diag(1, -1, 1)$, write
$$\tilde L(0)=\Phi^0diag(2, 1, -100)S^{-1}(\Phi^0)^TS, $$ where
\begin{eqnarray*}
\Phi^0=\left(
\begin{array}{ccc}
1 & -2 & 2 \\
2 & -3 & 2 \\
-2 & 2 &-1 \\
\end{array}
\right) \ .
\end{eqnarray*}
Calculating (\ref{DDD}), we get $D_1(t)=e^{4t}-4e^{2t}+4e^{-200t}$.
Since $D_1(0.4)=-3.95<0$, we conclude $\tilde L(t)$ blows up to infinity
before $t=0.4$.

\section{$\tau$-functions for the tridiagonal hierarchy}
\renewcommand{\theequation}{6.\arabic{equation}}\setcounter{equation}{0}

In this section, we only consider the case of $\tilde L$
being a tridiagonal matrix, and
change ${\bf t}$ to $2{\bf t}$ for convenience.
$\tilde L$ can be written as:
\begin{eqnarray}
\label{tildel0}
\tilde L = \left(
\begin{array}{lllll}
s_1a_1 & s_2b_1 & 0 & \ldots & 0 \\
s_1b_1 & s_2a_2 & s_3b_2 & \ldots & 0 \\
0 & \vdots & \ddots & \vdots & 0 \\
0 & \ldots & \ldots & s_{N-1}a_{N-1} & s_Nb_{N-1} \\
0 & \ldots & \ldots & s_{N-1}b_{N-1} & s_Na_N \\
\end{array}
\right) \ .
\end{eqnarray}
Similarly to the symmetric tridiagonal case, we can
introduce $\tau_i$'s as in (\ref{atau}). Equations
(\ref{atau}), (\ref{btau}) are now modified to,
\begin{eqnarray}
\label{matau}
s_ia_i=\frac{\partial}{\partial t_1}\log\frac{\tau_{i}}{\tau_{i-1}},
\end{eqnarray}
and
\begin{eqnarray}
\label{mbtau}
s_is_{i+1}b_i=\frac{\tau_{i+1}\tau_{i-1}}{\tau_i^2}.
\end{eqnarray}
However the equation for $\tau_i$  remains the same as (\ref{etau}).
The $\tau$-functions are also given by the symmetric wroskian (\ref{tau})
with $g=<(\phi_1^0(\lambda))^2e^{\xi(\lambda,{\bf t})}>$.  Note in particular
that
the function $g$ also satisfies (\ref{geq}) with the change ${\bf t}$ to
$2{\bf t}$, i.e.
\begin{eqnarray}
\label{geq2}
{\partial g \over \partial t_n} = {\partial^n g \over \partial t^n_1} \ .
\end{eqnarray}
In the symmetric tridiagonal case, it was shown in \cite{KM} that $\tau_i$'s
are given by $D_i$'s in (\ref{DDD}). In our case, $\tau_i$'s are related
to $D_i$'s by
\begin{eqnarray}
\label{tauD}
\tau_i=\frac{D_i}{s_1\cdots s_i} \ .
\end{eqnarray}
For instance, in example (i) of the previous section, $\tau_1$ is given
by $D_1$ in (\ref{d1}) and $\tau_2=-D_2=-\exp(-ct)$. It should be noted that in
the symmetric tridiagonal case, these
$\tau$-fuctions are positive definite, which is important
in the moment problem of Hamburger, while in our case,
the $\tau$-fuctions are no longer positive definite, in general.

\medskip

Let us now derive a hierarchy for $\tau_i$ in (\ref{tau}),
which includes (\ref{hirota}) as its first
member.  First we note the following:  In terms of $P_i(-\tilde \partial_{\bf
t})$,
 where $P_i$ are the elementary Schur polynomials defined by
\begin{eqnarray}
\label{generating}
e^{\xi(\lambda,{\bf t})} =\sum_{n=1}^{\infty}P_n({\bf t})\lambda^n,
\end{eqnarray}
(\ref{etau}) can be written as
\begin{eqnarray}
\label{ptau}
\left[ P_1^2(-\tilde \partial_{\bf t})\tau_i \right]\tau_i -
\left[ P_{1}(-\tilde \partial_{\bf t})\tau_i \right]^2=
\tau_{i-1}\tau_{i+1} \ .
\end{eqnarray}
where $\tilde \partial_{\bf t}=(\partial_{t_1},
\frac{1}{2}\partial_{t_2}, \cdots)$.
As a generalization of (\ref{ptau}), we have:
\begin{Lemma}
\label{ge}
For any $k, l\ge 1$ , the $\tau$-functions (\ref{tau}) satisfy
\begin{eqnarray}
\label{taues}
\nonumber
& \left[ P_k(-\tilde \partial_{\bf t})P_{l}(-\tilde
\partial_{\bf t})\tau_i \right] \tau_i -
\left[ P_k(-\tilde \partial_{\bf t})\tau_i \right] \left[ P_{l}(-\tilde
\partial_{\bf t})\tau_i \right] \\
& =
\left[ P_{k-1}(-\tilde \partial_{\bf t})P_{l-1}(-\tilde
\partial_{\bf t})\tau_{i-1} \right] \tau_{i+1}.
\end{eqnarray}
\end{Lemma}
\begin{Proof}
{}From the wronskian structure of the $\tau_i$ with $g$ satisfying
(\ref{geq2}),
we have
\begin{eqnarray}
\label{12}
P_k(-\tilde \partial_{\bf t})\tau_i=\tau_{i+1}(i-k+1, i+1),
\end{eqnarray}
where $\tau_{i+1}(i-k+1, i+1)$ is the determinant of $\tau_{i+1}$ after
removing the $i-k+1$th row and the $i+1$th column.  Note that
$P_k(-\tilde \partial_{\bf t})\tau_i=0$, for $k > i$.
For $1\le k,l\le i$,
we also get from the symmetric structure of the wronskian
\begin{eqnarray}
\label{23}
P_k(-\tilde \partial_{\bf t})P_{l}(-\tilde \partial_{\bf t})\tau_i
=\tau_{i+1}(i-k+1, i-l+1).
\end{eqnarray}
Noting $\tau_i = \tau_{i+1}(i+1,i+1)$ we have
\begin{eqnarray}
\label{rhs}
\nonumber
&\left[ P_k(-\tilde \partial_{\bf t}) P_{l}(-\tilde \partial_{\bf t})\tau_i
\right ]\tau_i - \left[ P_k(-\tilde \partial_{\bf t})\tau_i \right]
\left[ P_{l}(-\tilde \partial_{\bf t})\tau_i\right] \\
\nonumber
\ \ \ \ &=  \tau_{i+1}(i-k+1, i-l+1)\tau_{i+1}(i+1, i+1) \\
& - \tau_{i+1}(i-k+1, i+1)\tau_{i+1}(i+1, i-l+1).
\end{eqnarray}
The equation (\ref{taues}) then results directly from the the Jacobi formula
for a determinant $A (=\tau_{i+1})$,
\begin{eqnarray}
\label{jacobi}
A(i,k)A(j,l)-A(i,l)A(j,k)=A\left(
\begin{array}{cc}
i & k \\
j & l \\
\end{array}
\right) A,
\end{eqnarray}
where $A( \begin{array}{cc}
i & k \\
j & l \\
\end{array} )$ is the determinant of $A$ after removing the $i$th and $ j$th
 rows and the $k$th and $ l$th columns.
\end{Proof}

\medskip

We then multiply $\lambda^{-k}\mu^{-l}$ on the both sides of (\ref{taues}),
and sum up $k$ and $l$ to get, from (\ref{generating}),
\begin{eqnarray}
\label{exp}
\nonumber
&\left[\exp(-\sum_k\frac{1}{k\lambda^k}\partial_{t_k})
\exp(-\sum_{l}\frac{1}{l\mu^{l}}\partial_{t_{l}})\tau_i\right]\tau_i \\
\nonumber
& -\left[\exp(-\sum_k\frac{1}{k\lambda^k}\partial_{t_k})\tau_i \right]
\left[\exp(-\sum_{l}\frac{1}{l\mu^{l}}\partial_{t_{l}})\tau_i\right] \\
& = {1 \over \lambda\mu} \left[\exp(-\sum_k\frac{1}{k\lambda^k}\partial_{t_k})
\exp(-\sum_{l}\frac{1}{l\mu^{l}}\partial_{t_{l}})\tau_{i-1} \right]\tau_{i+1}.
\end{eqnarray}
Thus we obtain:
\begin{Proposition}
The $\tau$-functions satisfy a ``bilinear identity" generating the
relations of (\ref{taues}),
\begin{eqnarray}
\label{epsilon}
\nonumber
& \tau_i({\bf t}-\epsilon[\lambda]-\epsilon[\mu])\tau_i({\bf t})-
\tau_i({\bf t}-\epsilon[\lambda])\tau_i({\bf t}-\epsilon[\mu]) \\
& = \frac{1}{\lambda\mu}\tau_{i-1}({\bf t}-\epsilon[\lambda]-\epsilon[\mu])
\tau_{i+1}({\bf t}),
\end{eqnarray}
where $\epsilon[\lambda]=\left(\lambda^{-1}, \frac{1}{2}\lambda^{-2},
\cdots \right)$.
\end{Proposition}

\medskip

We can now obtain a hierarchy written in the Hirota bilinear
form (\ref{epsilon}). By setting $\mu=\lambda$, (\ref{epsilon}) reads
\begin{eqnarray}
\label{epsilon1}
\nonumber
& \tau_i({\bf t}-2\epsilon[\lambda])\tau_i({\bf t})-
\tau_i({\bf t}-\epsilon[\lambda])\tau_i({\bf t}-\epsilon[\lambda]) \\
& = {1 \over \lambda^2}\tau_{i-1}({\bf t}-
2\epsilon[\lambda])
\tau_{i+1}({\bf t}) \ .
\end{eqnarray}
Then changing ${\bf t}-\epsilon[\lambda]$ to {\bf t},
we can rewrite (\ref{epsilon1}) as
\begin{eqnarray}
\label{ab}
\nonumber
&\tau_i({\bf t}-\epsilon[\lambda])\tau_i({\bf t}+\epsilon[\lambda])-
\tau_i({\bf t})\tau_i({\bf t}) \\
& = \frac{1}{\lambda^2}\tau_{i-1}({\bf t}- \epsilon[\lambda])
\tau_{i+1}({\bf t}+\epsilon[\lambda]) \ .
\end{eqnarray}
This gives:
\begin{Corollary}
With the Hirota derivatives  $\tilde D_n :=D_{t_n}$ in (\ref{Derivative}),
we have
\begin{eqnarray}
\label{deo}
\left( {\exp(\tilde D) -1}\right) \tau_i\cdot\tau_i
=\frac{1}{\lambda^2}\exp(\tilde D)\tau_{i+1}\cdot\tau_{i-1} \ ,
\end{eqnarray}
where $\tilde D:=\sum_1^{\infty}{1 \over n\lambda^{n}}\tilde D_{n}$.
\end{Corollary}

Expanding (\ref{deo}) in the power of $\lambda^{-1}$, and  using
(\ref{generating}),
we obtain equations written in Hirota's bilinear form, for $n \ge 0$,
\begin{eqnarray}
\label{pd}
P_{n+2}(\tilde D)\tau_i\cdot\tau_i=P_{n}(\tilde D)\tau_{i+1}\cdot\tau_{i-1} \ .
\end{eqnarray}
The first three
equations in (\ref{pd}) are
\begin{eqnarray}
\label{ex1}
\tilde D_1^2\tau_i\cdot\tau_i=2\tau_{i+1}\cdot\tau_{i-1} \ ,
\end{eqnarray}
\begin{eqnarray}
\label{ex2}
(\tilde D_1\tilde D_2)\tau_i\cdot\tau_i=2\tilde D_1\tau_{i+1}\cdot\tau_{i-1} \
,
\end{eqnarray}
\begin{eqnarray}
\label{ex3}
(\tilde D_1^4+3\tilde D_2^2+8\tilde D_1\tilde D_3)\tau_i\cdot\tau_i=
12(\tilde D_2+\tilde D_1^2)
\tau_{i+1}\cdot\tau_{i-1} \ .
\end{eqnarray}
Note here that (\ref{ex1}) is just (\ref{hirota}) with ${\bf t}$ changes to
$2{\bf t}$, and the ``odd'' powers of $\tilde D_i$'s on the l.h.s. of
(\ref{pd})
are suppressed by operating on $\tau_i\cdot\tau_i$.

\par\medskip\medskip

{\bf Acknowledgment}
The work of Y. Kodama is partially supported by an NSF grant DMS9403597.

J. Ye wishes to thank Prof. Guangtian Song for his encouragement through the
years.


\bibliographystyle{amsplain}

\end{document}